\documentclass[sigconf]{nimeart}

\usepackage{xurl}
\usepackage{CJKutf8}
\setcopyright{cc}
\copyrightyear{2026}
\acmYear{2026}
\acmDOI{}
\acmConference[NIME '26]{International Conference on New Interfaces for Musical Expression}{June 23--26,
  2026}{London, UK}
\acmISBN{}

\begin{document}

\title{Opening the Design Space: Two Years of Performance with Intelligent Musical Instruments}

\author{Charles Patrick Martin}
\email{charles.martin@anu.edu.au}
\orcid{0000-0001-5683-7529}
\affiliation{%
  \institution{The Australian National University}
  \city{Canberra}
  \state{ACT}
  \country{Australia}
}


\begin{abstract}
Machine generation of symbolic music and digital audio are hot topics but there have been relatively few digital musical instruments that integrate generative AI.
Present musical AI tools are not artist centred and do not support experimentation or integrating into musical instruments or practices.
This work introduces an inexpensive generative AI instrument platform based on a single board computer that connects via MIDI to other musical devices.
The platform uses artist-collected datasets with models trained on a regular computer.
This paper asks what the design space of intelligent musical instruments might look like when accessible and portable AI systems are available for artistic exploration.
I contribute five examples of instruments created and tested through a two-year first-person artistic research process.
These show that (re)mapping can replace retraining for discovering AI interaction, that fast input interleaving is a new co-creative strategy, that small-data AI models can be a transportable design resource, and that cheap hardware can lower barriers to inclusion.
This work could enable artists to explore new interaction and performance schemes with intelligent musical instruments.
\end{abstract}


\maketitle

\section{Introduction}\label{intro}


\begin{figure}[t]
    \centering
    \includegraphics[width=0.95\linewidth]{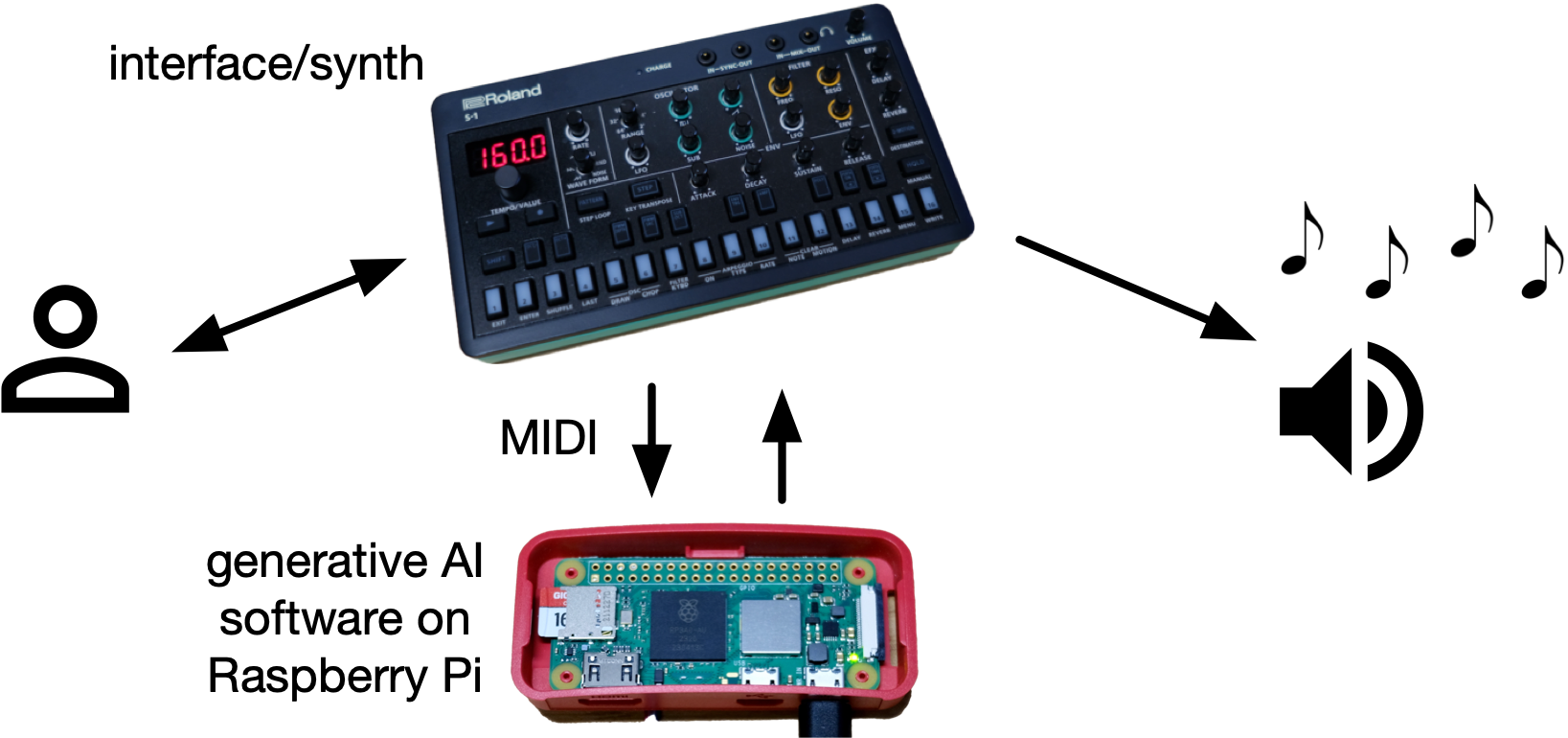}
    \caption{
    The generative AI interactive music platform with a hardware synthesiser. 
    This platform connects to other devices via MIDI, receiving signals from a human performer and send signals to control sound on the synthesiser. 
    The software is pre-installed on a Raspberry Pi operating system image and runs on even the cheapest Raspberry Pi Zero 2 W (15USD).}
    \label{fig:system-diagram}
\end{figure}

This paper explores how generative AI may be embedded within interactive musical systems for live performance through a series of case study instruments and performances where existing hardware and software instruments have been coupled with a Linux-based single board computer pre-loaded with generative AI software.
A broad range of music technology research has applied AI techniques, and in particular deep learning, for tasks such as mapping, audio or sensor analysis and musical data generation~\cite{jourdan_nime_ml_review}. 
While the analysis of gestures~\cite{visi_iml_gesture} remains a popular application, deep learning models now enable generation of audio~\cite{caillon_rave}, symbolic music~\cite{ji_survey_2024}, and interaction data~\cite{Martin2019}.
Despite this wide range of research, relatively few intelligent musical instruments, that is, digital musical instruments (DMIs) integrating generative AI, are available to be used in ordinary musical practice.
The goal of this research is to expand the space of intelligent musical instrument design by prototyping and performing with multiple new instruments, examining them through artistic research, leading to actionable design insights.
The approach is to introduce a platform (IMPSY)~\cite{impsy_software_zenodo} for creating intelligent instruments that is cheap, small and battery-powered, thus enabling integration into a wider variety of musical setups and practices across a varied community of artists and experimenters. 
The new platform includes AI software for generating expressive musical signals from small artist-centred datasets, runs on inexpensive Raspberry Pi computers, and is configurable via a web interface.
As an initial step towards identifying a wider variety of intelligent musical instrument designs, this paper reports a series of five instruments and performance experiences from a first-person perspective, illustrating different ways that this system can be incorporated into an artistic practice.

AI techniques have been applied in music throughout the history of computing~\cite{roads-music-AI-1985}. Real-time interactive systems became more prominent from the 1990s such as Continuator~\cite{Pachet:2003wd},  Voyager~\cite{Lewis:2000fu}, GenJam~\cite{Biles:2007aa} as well as tools such as Wekinator~\cite{fiebrink_meta-instrument_2009} for embedding machine learning models in new musical systems.
New and more capable deep learning techniques have led to an expansion of interest in AI music making at CHI workshops~\cite{genaichi2022}, music industry white-papers~\cite{goldmedia_ai_2024}, and the general media~\cite{musicpublishers}. Generative AI is now applied in tasks such as sound design~\cite{sound_design_ai_2024}, symbolic music generation~\cite{ji_survey_2024}, and music production~\cite{deruty_development_2022}. Its use in live musical performances is still only beginning to be explored through NIME and related communities through practice-based methodologies~\cite[e.g.,][]{privato-hauntography-2024, stefansdottir_intelligent_2025}.
With many now framing generative AI as a threat to musicians' livelihoods~\cite{goldmedia_ai_2024} and fraught with ethical pitfalls~\cite{ethical_genaudio_2023}, it is appropriate to introduce and study artist-centred approaches to creating music with AI.

This work introduces a small-data~\cite{Vigliensoni:2022} approach to embedding AI within musical instruments where artists collect, curate, train, and deploy their own intelligent musical instruments.
I define an intelligent musical instrument as an instrument where an AI system generates actions independently of a musician's actions.
This definition would encompass Continuator~\cite{Pachet:2003wd} and Voyager~\cite{Lewis:2000fu} but exclude systems such as Wekinator~\cite{fiebrink_meta-instrument_2009} where AI is used for mapping musicians' actions or interpreting sensor data but not generating independent actions.
The platform introduced in this paper allows a generative AI algorithm to be incorporated within existing electronic music devices through standard MIDI communication. 
The AI algorithm is able to control not just notes but the significant variety of timbral parameters that are available in electronic instruments. 
Similarly, the AI platform can respond to the musician's actions on standard interfaces such as piano-style keyboards as well as other controls such as knobs, sliders, levers, or movement sensors that are typically part of electronic music devices (e.g., see Figure \ref{fig:system-diagram}).
The AI platform's interaction mappings are configured through a web interface which is also used to retrieve recorded data for re-training the system's machine learning model.

With this platform, intelligent musical instruments can be quickly prototyped and tested through musical performance.
Incorporating generative AI within electronic instruments leverages musicians' existing significant skills and knowledge which could enable deeper co-creative designs.
In this work, I describe the new AI music platform, showing that it is affordable and achievable for non-experts in AI to use.
I document five intelligent musical instruments (the Intelligent Volca, MicroFreak, S-1, DAW, and Setup) and reflections on two years of performance practice with this system.
The results contribute an expansion of the design space for intelligent musical instruments, showing that (re)mapping can replace retraining for discovering AI interaction, that fast interleaving between human and AI input is a new co-creative strategy, that small-data AI models can be a transportable design resource, and that cheap hardware can lower barriers to inclusion.
More broadly, it has been argued that explorations with technology within artistic practices can contribute to HCI~\cite{improvisation-arts-hci-2018, artistic-narratives-hci, state-of-arts-inchi-2022, interactive-art-hci-2019}.
This work uses artistic methodologies to contribute to the broader conversation on design principles and patterns for generative AI in HCI~\cite{design-principles-genai-2024} by examining design possibilities that might encourage more widespread prototyping, hacking, and musicking with AI in musical performance.

\section{Related Work}\label{related-work}


Jourdan and Caramiaux identified gaps in interactivity and practice~\cite{jourdan_nime_ml_review} in AI-supported musical expression, particularly where deep learning is applied. 
They argued that as deep learning systems have become more complex there have been fewer examples of application in long-term musical practice and artists have less interaction with data collection and training phases. 
This contrasts with early AI music systems such as Voyager~\cite{Lewis:2000fu} that were developed over multiple decades and parameter-mapping approaches where long term practice has been studied~\cite{fiebrink_years_ml}.
Tahiroğlu's AI-terity instrument~\cite{tahiroglu-aiterity-2021}, a custom physical interface for exploring AI sound generation while also acting autonomously to engage the performer, has undergone  artist-centred evaluation. 
The experience of performing with AI-terity was later analysed to understand how the system prompted ``unfamiliar musical expectations''~\cite{tahirouglu-evolving-2022}.
This role mirrors some of Parkinson and Dunning's findings for roles that musical machines, including intelligent musical instruments, might have~\cite{taxonomy-music-interactions-parkinson-2025}. They argue that such instruments can ``evolve'', ``collaborate'', and ``learn'' (among other roles). Generative AI may be well placed to help enhance these roles within new musical instruments. 
Recent practice-based approaches to intelligent instrument study have articulated how small datasets can be curated and applied with care and sensitivity~\cite{stefansdottir_intelligent_2025}.
Generative AI algorithms have also been embedded within software-based instruments. For example, Magenta Studio~\cite{magenta-studio-2019} seeks to embed generative AI algorithms within the Ableton Live digital audio workstation (DAW) software as a plugin that can generate and manipulate MIDI recordings. While these have been examined from an artist perspective~\cite{Davis:2022}, the focus is music production or composition not live performance.


Incorporating computer music systems into embedded hardware systems can allow better integration into artistic practices with many recent examples using Raspberry Pi~\cite{Berdahl:2013aa} and Beaglebone~\cite{mcpherson-bela-2016} single-board computers. 
Research applying generative AI on such single board computers promises to allow these technologies to be embedded within self-contained musical instruments. 
This has been demonstrated for Raspberry Pi~\cite{Naess:2019aa, martin_empi} and Bela~\cite{pelinski_pipeline_bela} platforms. 
Previous work has acknowledged the difficulties faced by instruments builders wishing to embed AI within new instruments in terms of data collection and training~\cite{Martin2019} and cross-compiling code~\cite{pelinski_pipeline_bela}. These efforts are mirrored in other creative design fields~\cite{teachablemachine}.
Pelinski et al.~\cite{pelinski_pipeline_bela} demonstrated a pipeline for collecting data and training machine learning models for musical instruments built using the Bela embedded platform~\cite{mcpherson-bela-2016}.
Martin et al.~\cite{martin_empi} examined an embedded AI-based musical instrument and how different embodied representations of musical actions were perceived by improvisers.
In the Sophtar, a custom string instrument with configurable interface components and embedded computer, Visi focuses on the evolving design process of a physical instrument with AI sound modelling and autonomous interactions~\cite{visi-sophtar-2024}.
In contrast with very large and costly industrial-scale AI, these projects have emphasised a small data~\cite{Vigliensoni:2022} mindset where artists should collect, curate, train, and deploy their own AI systems. 


This research explores generative AI instruments from a first-person or autoethnographic perspective~\cite{autoethnography-hci} where instruments are created and tested through autobiographical design~\cite{autobiographical-design}. 
This is related to the practice-based or -led approach~\cite{Smith:2009gf} and artistic research methods~\cite[e.g.,][]{bowers-room-feedback} often applied within creative arts research where knowledge is distilled from an artistic process. Artistic processes have been previously applied as sources of HCI knowledge~\cite{artistic-narratives-hci}, for understanding intelligent instrument practice~\cite{privato-hauntography-2024, stefansdottir_intelligent_2025}, and for expanding instrument design spaces~\cite{bowers-room-feedback}. As the goal of this research is to expose design possibilities, it is appropriate to apply a first-person artistic perspective as an initial step with the ultimate goal of incorporating a wider range of perspectives through artist-centred methods.

\section{System Design}\label{sys-design}

\begin{figure*}
    \centering
    \includegraphics[width=0.45\linewidth]{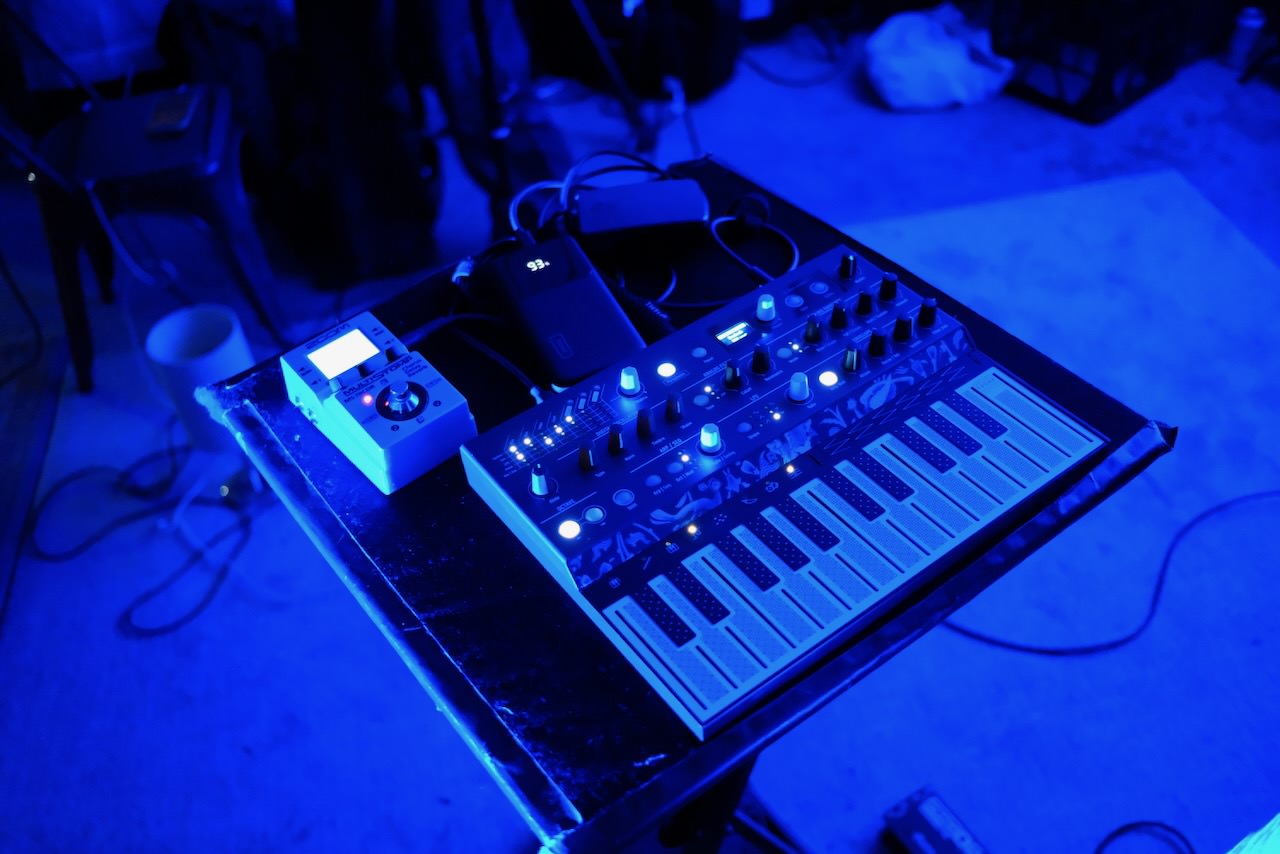}
    \includegraphics[width=0.35\linewidth]{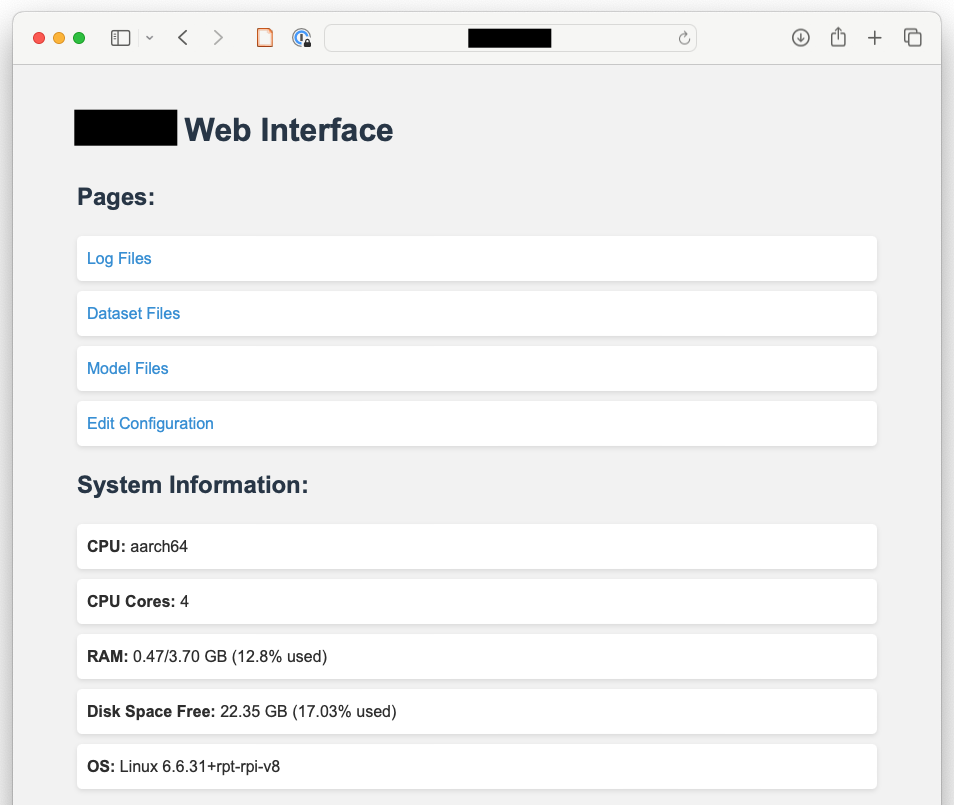}
    \caption{The stage setup for an intelligent musical instrument using the generative AI platform running on a Raspberry Pi 4 (left) and the web interface for configuration (right). The complete setup of synthesiser, Raspberry Pi, and effects pedal is battery powered. The software reacts to the musicians' actions through knobs and keys on the synthesiser front panel and controls notes and timbral parameters. Video examples are available here: \url{https://doi.org/10.5281/zenodo.19550146}}
    \label{fig:impsy-platform-interface}
\end{figure*}

The generative AI interactive music platform consists of Python software, a Raspberry Pi single board computer, and a custom operating system image that communicates to external electronic music devices over a USB, MIDI, or network connection. 
This setup is not intended to produce sound but only to control other electronic music devices and so the single board computer does not need to have an audio output. 
The platform is designed so that a musician can install the operating system image, which runs the AI software on boot, onto an SD card and then configure the platform from their computer using a web browser. 
After an initial setup the platform can be incorporated into an electronic music setup for performances and experiments without further configuration. 
If desired, data recorded to the SD card can be retrieved and used to train new AI models for the system.
A diagram of the platform in use and the web interface is shown in Figure \ref{fig:impsy-platform-interface} and a system diagram for a typical setup is shown in Figure \ref{fig:system-diagram}.

\subsection{Hardware}

The platform uses the Raspberry Pi family of single board computers which are widely available and familiar for many electronic instrument designers. 
The software works on all Raspberry Pi models that support the 64-bit Raspberry Pi OS which ensures good compatibility with machine learning libraries for Python. 
This includes the inexpensive Raspberry Pi Zero 2 W (15USD as of 2026) that features a quad-core 64-bit ARM Cortex-A53 at 1GHz with 512MB of RAM and which is small enough (65mm x 30mm) to be incorporated inside new instrument designs.

\begin{figure}[t]
    \centering
    \includegraphics[width=0.95\linewidth]{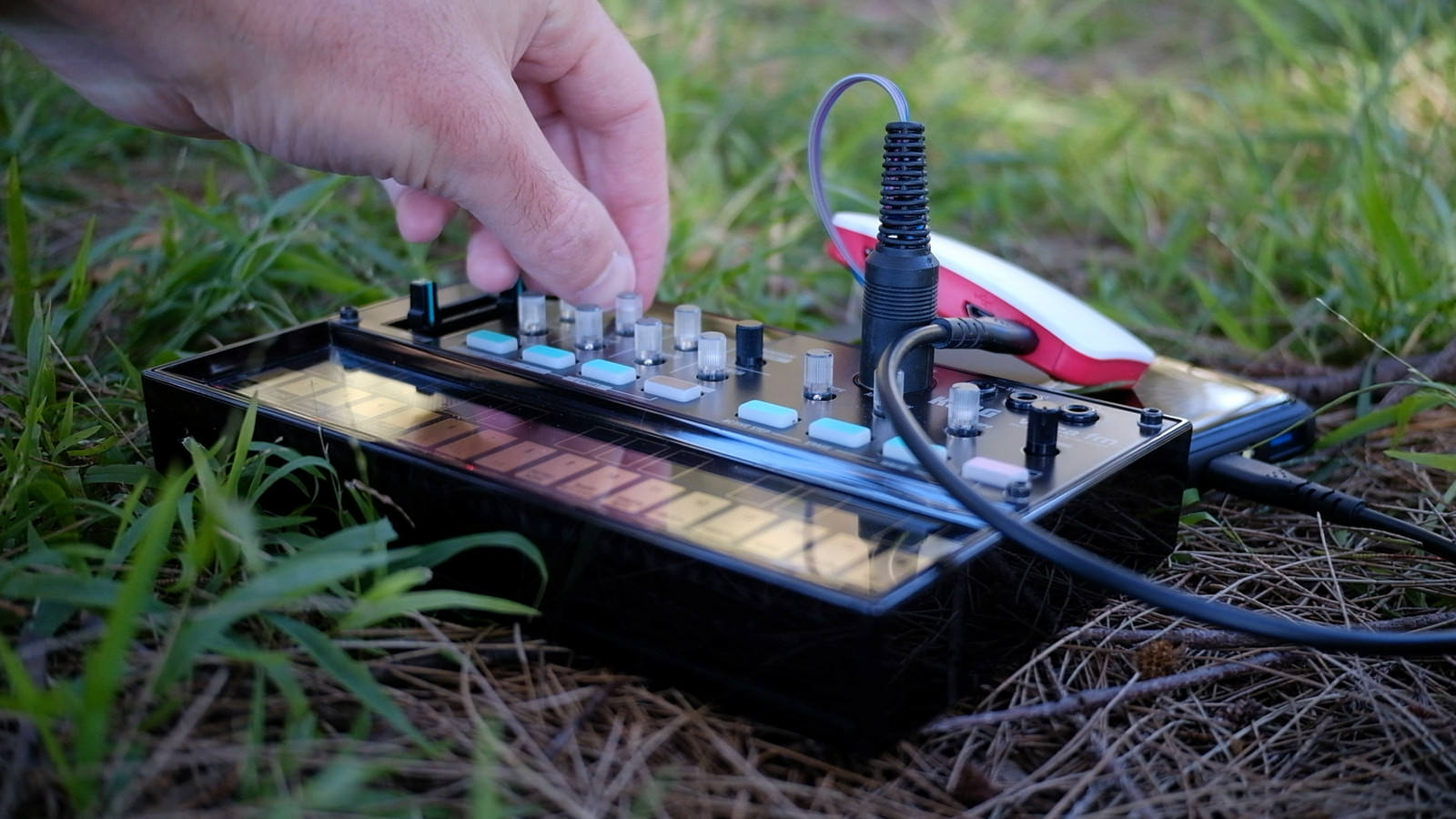}
    \caption{Battery-powered human-AI interaction on a Korg Volca FM. The generative AI system running on the Raspberry Pi Zero controls pitch and rhythm via the MIDI connector while the musician edits the synth patch. 
    }
    \label{fig:self-playing-volca}
\end{figure}

Several methods of communicating with other devices from the Raspberry Pi are possible.
MIDI is used as the primary communication format and the software can send MIDI messages via either a USB-connected MIDI interface, a direct USB MIDI connection to a hardware synthesiser, or the Raspberry Pi's serial (UART) output. 
The software can also communicate over a network using OSC (open sound control) or WebSockets messages. 
The smallest Raspberry Pi Zero models only support one USB host connection while larger models (e.g., 4 Model B or 5) have multiple USB host ports and an Ethernet interface so could support multiple communications channels simultaneously.
For MIDI-over-serial connections, the MIDI connector can be soldered to the GPIO pads corresponding to the Pi's serial (UART) output. 
Providing a MIDI output from the Pi's 3.3V signals is straightforward requiring only two resistors~\cite{MIDI-Manufacturers-Association:2014aa}: 10$\Omega$ between the UART TX pin and pin 5 on the MIDI connector and 33$\Omega$ between the 3.3V output and MIDI pin 4. 
MIDI pin 2 is connected to ground. 
In the setup shown in Figure \ref{fig:self-playing-volca}, the resistors are concealed within the MIDI connector.

\subsection{AI model}

Following Martin and Torresen~\cite{Martin2019}, the new system uses a mixture density recurrent neural network (MDRNN) and the Keras-MDN-Layer library~\cite{keras-mdn-layer}.
This machine learning paradigm can generate a stream of musical data in free rhythm that models embodied musical gestures meeting the intention of controlling the parameters of electronic musical instruments in improvised musical performances.
The system uses an MDRNN that generates tuples of data that are interpreted as number of musical values and a time delta for when that value should occur in the future. 
The number of musical values is configurable within the software but is at a minimum one, i.e., the system models one musical parameter over time. 
The neural network is autoregressive, so takes the preceding value as input to generate the next, but it is also recurrent, so it stores a lossy history of generated values using LSTM units. 
In this research, small MDRNN models (e.g., 2 layers of 64 LSTM units) are generally used to model 1--8 parameters.
Although these are very small compared to typical generative AI and large language models, they are sufficient for generating musically useful data in real-time on small hardware.
The time for training a new model depends on the size of the training data and speed of the training computer, but effective models can be trained in under 30 minutes on a normal laptop.


\subsection{Software}

The system's software is a Python program running on the Raspberry Pi that generates musical data and sends it to external electronic musical instruments.
This program listens to MIDI signals from the instrument and responds with AI-generated continuations of the performance state.
MIDI note-on as well as control change messages are supported so the system is able to play notes as well as change the timbre of a connected instrument.
As mentioned above, the AI model is capable of listening and responding to a number of channels of MIDI data simultaneously.
The software records all data that is received from the MIDI interface and stores this as timestamped logs. 
This means that performing with this system builds up new datasets allowing new AI models to be trained.

An initial configuration is required to choose the MIDI interface and the specific MIDI signals to listen for and send out. 
Configuring interaction mappings is a design task that fundamentally changes how a created intelligent musical instrument will function. 
For example, it may be interesting to only assign timbral changes to the generative AI system while a performer controls notes (or vice-versa). 
Alternatively, a performer's notes can be cross-mapped to influence the timbral parameters changed by the generative AI platform, or the input and output of the AI model mapped to different devices.
A simple web interface is provided for adjusting the configuration file, downloading logged data of previous performances and uploading new AI models.
The configuration capability for the system evolved over the two-year time span considered in this paper, as described in Section \ref{sec:case-studies}.

The software for the system can be installed using Poetry or Python's pip tool but this is a laborious process for new users. 
To accelerate setup with a new instrument, software is pre-installed on a Raspberry Pi OS image that can be flashed to an SD card. 
The generative AI interaction system and web server programs are started on boot and Ethernet-over-USB is enabled so that the Raspberry Pi can be directly connected to a computer for configuration.
The software is open source~\cite{impsy_software_zenodo} and available online\footnote{Source: \url{https://github.com/cpmpercussion/impsy}, Raspberry Pi images: \url{https://github.com/cpmpercussion/impsy-pi}}.

\subsection{Benchmarks and speed testing}

\begin{figure}[t]
    \centering
    \includegraphics[width=0.99\linewidth]{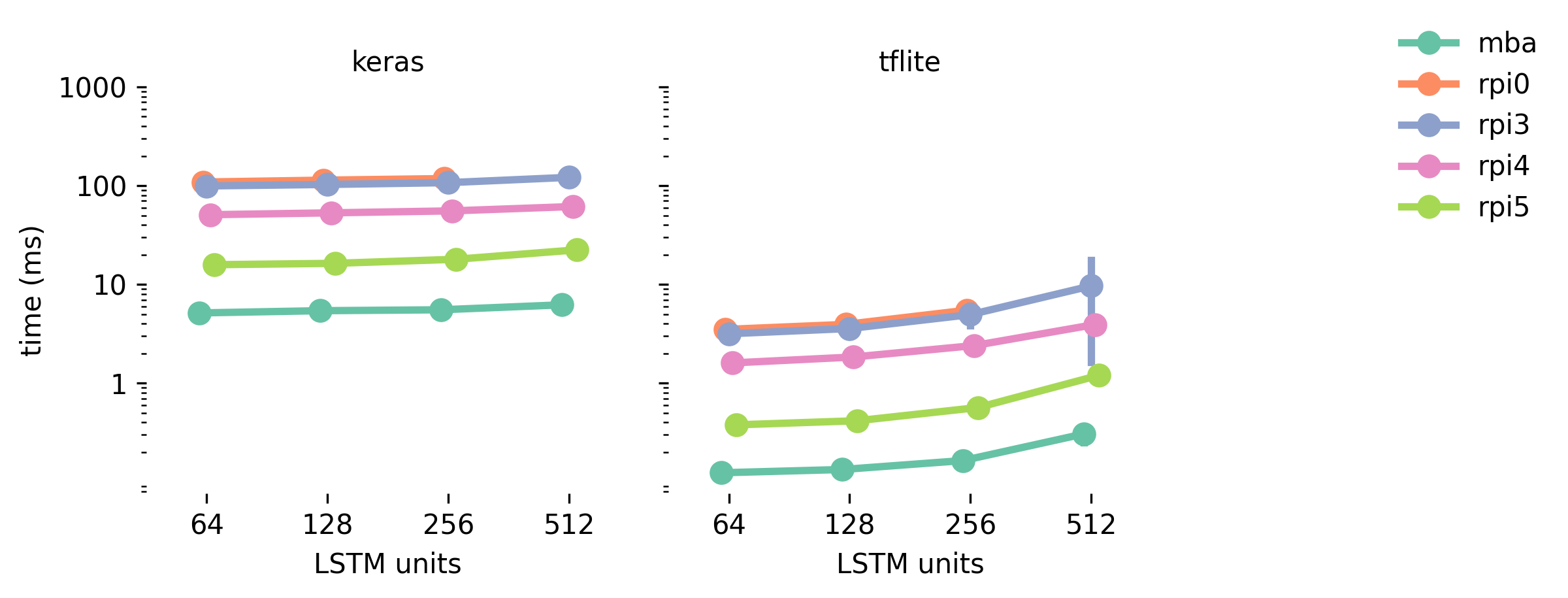}
    \caption{Inference time for differently sized AI models on Raspberry Pis and an Apple MacBook Air (M1, 16GB). All Pis can run AI predictions in $<5\textrm{ms}$.}
    \label{fig:inference-time}
\end{figure}

\begin{table}[t]
\caption{Boot times for different Raspberry Pi models measured from power-on to first MIDI output from the AI model.}
\label{tab:boot-times}
\begin{tabular}{@{}ll@{}}
Raspberry Pi Model and RAM & Power-to-Sound (s) \\ \midrule
Zero 2 W (512MB)   & 114                \\
4 B (2GB)               & 78               \\
5 (4GB)               & 38              \\
\end{tabular}
\end{table}

While quantitative testing is not a main focus of this paper, MDRNN inference speed testing previously reported in \cite{Martin2019} was replicated to compare saved model formats and Raspberry Pi versions. 
The goal was to establish expectations for different system configurations in creative application compared with a normal laptop computer.
Model formats compared were the Keras library's native format (keras) and the optimised Tensorflow Lite format (tflite). The tflite format was faster on all platforms and even the cheapest Raspberry Pi Zero 2 W runs AI predictions in $<5\textrm{ms}$ (Figure~\ref{fig:inference-time}), below the previous benchmark of 10ms~\cite{Martin2019}. 
For smaller models, the Raspberry Pi 5 runs AI predictions in less than 0.5ms.
These tests revealed that while the Pi Zero 2 W had useful performance on smaller neural networks (considering number of LSTM units as a measure of size), it ran out of memory and did not complete tests on the larger 512-unit model.
The boot-up time for different Raspberry Pis is another important aspect for integration in musical setups. 
The Zero 2 W was slow to boot at 114s while the Raspberry Pi 5 starts playing in 38s (Table~\ref{tab:boot-times}).

\section{Performance Experiences}\label{sec:case-studies}

This section describes the development of five new intelligent musical instruments using the generative AI platform and the experience of using them in performance over a period of approximately two years (2024-2026). 
These instruments were the Intelligent Volca, MicroFreak, S-1, DAW, and Setup. 
In this autobiographical artistic research, these instruments were developed for my own performance practice and tested in live gigs and recordings to expand knowledge about the design space for intelligent musical instruments. 
The performance style was free improvised music in solo, duo, and group situations.
These performance experiences include solo recordings and concerts as well as applying intelligent instruments within ensembles of acoustic and electronic instruments. 
I introduce the instruments in roughly chronological order, reflecting on how they were used in performance and how the interaction mappings and creative approaches evolved over time. 
These reflections cover the 15 performances and recordings listed in Table~\ref{tab:perf-corpus}. A demo video with each system is available as a supplementary media file with this paper and online\footnote{Video examples can be found here: \url{https://doi.org/10.5281/zenodo.19550146}}.

\subsection{The Intelligent Volca}

The Intelligent Volca was a proof-of-concept to establish whether real-time generative-AI MIDI signals from the Raspberry Pi Zero was feasible and interesting.
MIDI output was obtained from the Raspberry Pi's GPIO header UART pins and connected to the battery-powered Volca FM.
The AI model was trained on a corpus of around one hour of musical human interaction with a single continuous controller.
The AI model output (0.0--1.0) was converted to a MIDI pitch value from 0--127 and a corresponding MIDI note-on message scheduled to be sent at the time-delta generated, so the model could control pitch and rhythm on the Volca.

This initial experiment emphasised portability with a completely self-contained and battery-powered system; the Volca FM even has an internal speaker.
The generative AI system played notes while I could adjust the synthesiser's timbre or play overlapping notes on the keyboard; however, the AI model couldn't respond to me as the Volca FM has no MIDI output.
This meant that although this experiment started to explore different roles for a human and AI system in controlling the synth, the interaction was one way. 
The AI model, trained on expressive interaction with a continuous controller, tended to perform glissandi of various speeds, starting points and lengths on the Volca.
These sounded somewhat unusual as the synthesiser re-triggered its envelope for each note.
After recording a demo video with the system, I felt that the AI model might be more applicable controlling synthesis parameters where a musician would like a particular value to change smoothly in an expressive way.

\subsection{The Intelligent MicroFreak and S-1}

\begin{figure*}
	\centering
	\includegraphics[width=0.49\linewidth]{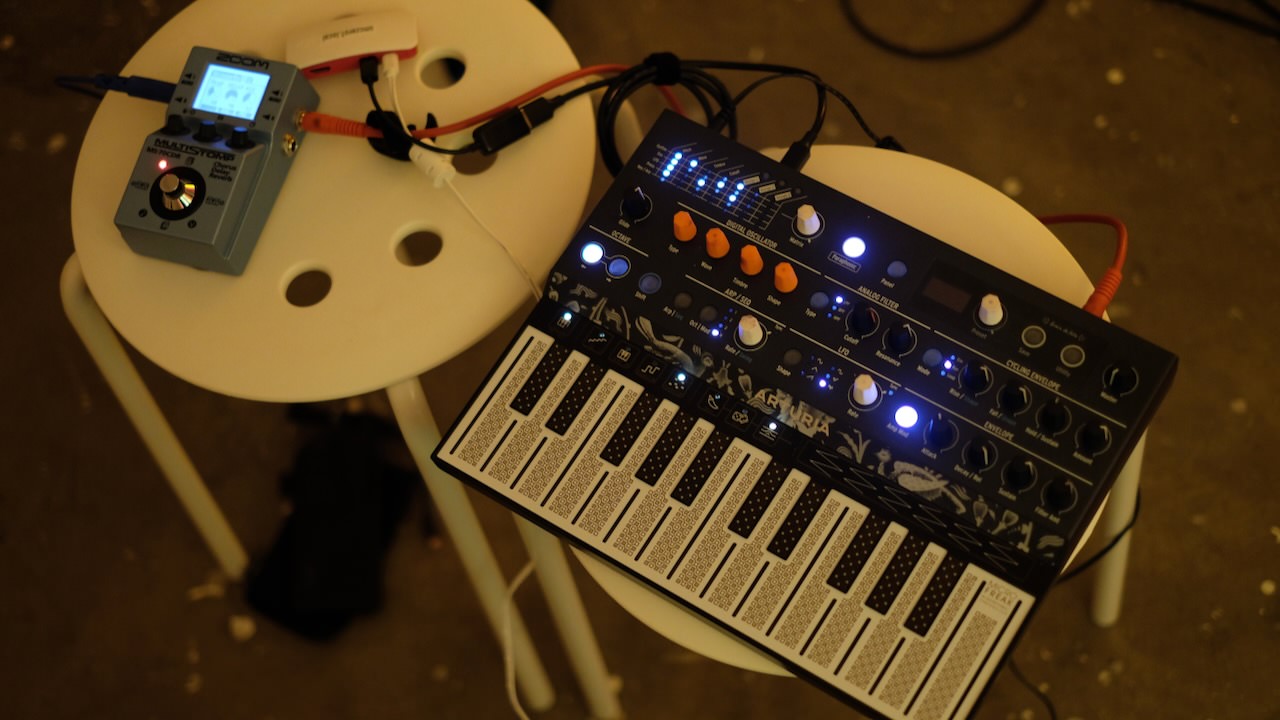}
    \includegraphics[width=0.49\linewidth]{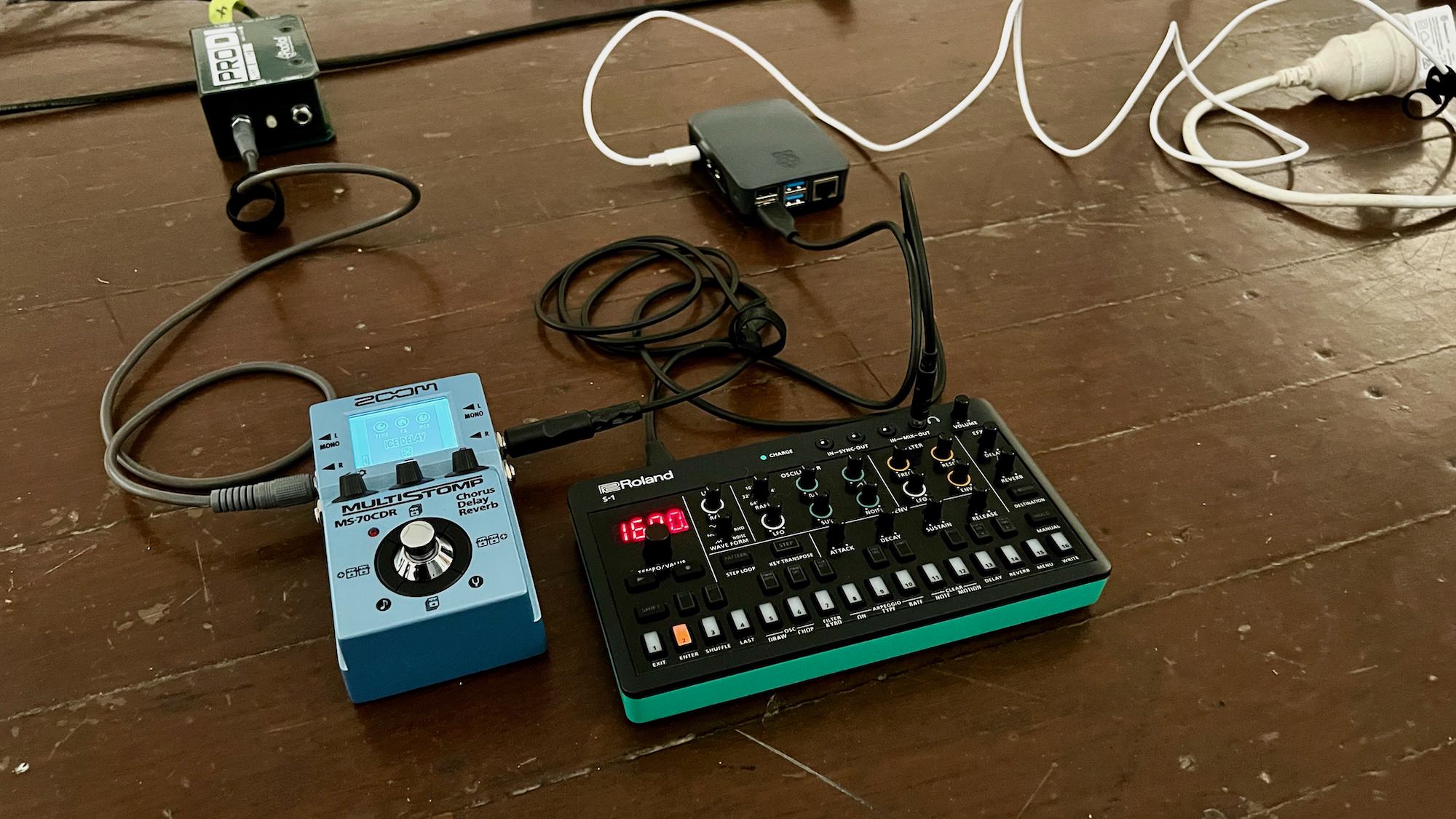}
	\caption{Stage setups with an Arturia MicroFreak synthesiser (left) and Roland S-1 (right).
    The software ran on a Raspberry Pi Zero 2 W (left) or Raspberry Pi 4 (right) and in both cases a Zoom effects pedal provided additional audio effects.
    For both setups, the generative AI platform generated note data (tracking the keyboard) and seven timbral parameters (tracking knobs on the synthesiser interface).}
	\label{fig:intelligent-synth}
\end{figure*}

Hardware synths with MIDI outputs and inputs promised to support two-way human-AI interaction.
The capacity for MIDI-over-USB connections was developed so the system could connect directly to synthesisers with built-in USB-MIDI interfaces where notes from their keyboard and control changes from their knobs and other interface elements can be sent and received over MIDI.
In many cases, simultaneous control of the synthesiser from the hardware interface and from MIDI signals is possible allowing shared human-AI interaction. 
The interaction loop for the AI system responded in a call-and-response manner so that when the performer adjusted the synth controls and played notes, the AI system tracked these signals. When the human stopped for a certain amount of time the AI system took over the controls.
My intention was to explore what music could be made by sharing control of the synths with this AI platform. 

I first used an Arturia MicroFreak synthesiser with the generative AI system controlling eight parameters: note-on messages, and seven timbral parameters available to the performer via knobs on the synthesiser face.
This instrument was used in individual and group improvisations with performers on a variety of other instruments.
The instrument was portable (see Figure \ref{fig:intelligent-synth}) with the tiny Raspberry Pi Zero and an effect pedal accompanying the small synthesiser.
This AI model was trained on data sourced from my improvisation on eight continuous controllers.
In performance, the AI model produced musically satisfying material.
Unlike a human, the AI system is able to adjust many parameters simultaneously, resulting in an inhuman but exciting exploration of the synth parameter space.
In non-AI performances, I would primarily play notes on the synthesiser with occasional timbral changes; however, the AI interactions update the timbral parameters in between phrases and in some cases in between notes, leading to varying and unique sounds. 
To enhance this effect, I tried setting the call-and-response switch-over time to a (very short) 0.1 seconds.

This performance concept was further explored with a Roland S-1 synthesiser where a similar mix of note and parameters signals were controlled by the AI system.
The very small size of this synth compromises the playability of the keyboard but using the AI system led to less reliance on the keys. 
In performance, I focussed more on timbral parameter setting, allowing the AI system to improvise notes in between human-selected parameter changes which were also under the control of the AI model leading to surprising timbral outcomes.

\subsection{The Intelligent Digital Audio Workstation}

\begin{figure}[t]
    \centering
    \includegraphics[width=0.99\linewidth]{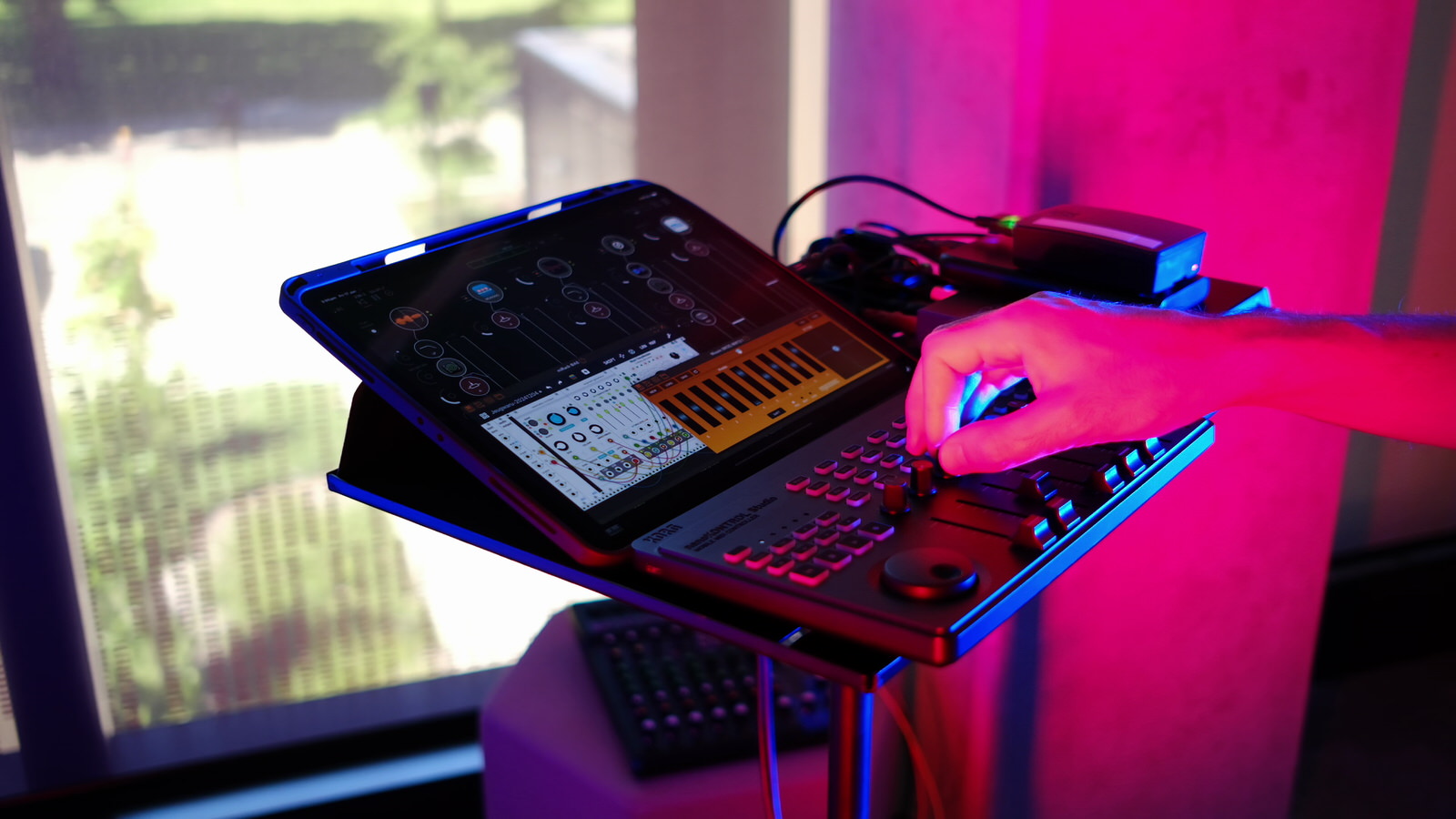}
    \caption{Software synthesisers used in an intelligent instrument setup. Digital audio workstation (DAW) software ran on the iPad which is connected via MIDI to the generative AI system on a Raspberry Pi 4. A performer can interact with the instrument using the iPad touchscreen or a hardware controller interface. 
    }
    \label{fig:intelligent-daw}
\end{figure}

Much contemporary electronic music is performed on software synthesisers and audio files loaded into digital audio workstation (DAW) software~\cite{daws-reuter-2021, Clauhs:2023aa}.
DAW software supports MIDI input and output to control parameters with internal routing and mapping possibilities.
I experimented with connecting the system to a DAW running on a computer or tablet via an inexpensive USB MIDI interface.
This represents a convenient way to design new intelligent musical instruments and interactions as both the DAW and generative AI system are highly configurable.

An example is illustrated in Figure \ref{fig:intelligent-daw} where a DAW and software plugin host, Kymatica AUM (Audio Mixer), ran on an iPad with eight different sound generators available in channel strips with associated effects.
The sound generators were a mix of software synthesisers and audio file players.
The generative AI system was configured to respond to eight different MIDI control change channels and generate four different channels of MIDI notes and four channels of control change data.
This input and output data is routed internally in AUM and assigned to control different parameters and notes within the sound generators.
I could influence the AI generated data on the iPad touchscreen via a software MIDI controller or with a hardware MIDI controller connected to AUM with Bluetooth.
This setup is very flexible in that different sound generators can be loaded or unloaded and the MIDI routing can be adjusted for different performance requirements without necessarily reconfiguring the generative AI software on the Raspberry Pi.

Mapping generative AI signals over a simple MIDI connection is an advantage for performing with tablets or laptops where installing a Python AI application is either impossible or inconvenient.
I used this specific setup in solo and ensemble electronic music improvisations.
The ability to re-map MIDI signals from the generative AI system to any parameter in the software synths was a way to evolve the intelligent instrument without retraining or modifying the generative AI system. 
I could map AI signals to areas where they were musically useful and swap software synths out with others or adjust them to sound significantly different. 
Having multiple synths loaded in a DAW allowed the output sound to be mixed and adjusted leading to quite different musical characteristics in between performances. 
While all parameters could be adjusted within the touchscreen interface, adding the hardware MIDI controller with physical knobs for influencing the generative AI system let me keep track of the autonomous components of the instrument.

\subsection{Intelligent Setups}

The above experiments and experiences focussed on a simple use-case where one synthesiser or computer was used for sound output as well as interface input; however, a natural extension is to support multiple interfaces simultaneously, each potentially capable of control input and sound output. 
In this kind of \emph{intelligent setup}, the AI model mediates events in between the different interfaces. 
The intelligent DAW idea explained above could enable this kind of extension with MIDI routings within the DAW itself, but a better solution for a DAW-less setup was to extend the MIDI mapping capability of the AI system to enable multiple input and output routings.

This idea was originally motivated to support an additional controller for the Roland S-1 synth which has very small control knobs and no visible feedback for synth parameters. 
A first experiment with the intelligent setup combined the S-1 with a Behringer X-Touch Mini controller with 8 (larger) knobs with LED rings to provide feedback.
The new mapping enabled the 8 knobs to control the AI model with human control events mapped directly through the S-1. 
Similarly, control on the S-1 was still available. 
MIDI outputs to control change could be restricted to a specific output range enabling control over the LED rings.
This first setup worked well in performance, but the new flexibility of the MIDI mapping encouraged experimentation with interfaces beyond knobs. 
Substituting the X-Touch for a Keith McMillen QuNeo pad controller enabled control of and feedback from the 8 AI-linked parameters through touch sliders and additional note control with the drum pads.

In practice, the intelligent setup idea allowed more freedom in choosing how I could control an intelligent instrument and what a performance might look like. 
With the QuNeo and S-1, I could spend bigger chunks of performances playing (by, e.g., triggering lots of notes from the drum pads), effectively stopping AI control of the S-1.
The QuNeo control strips made it easier to select reasonably specific values of multiple parameters quickly while the S-1 was more useful for slow sweeps through parameters.
With this setup, availability of different control gestures led to different ways of interacting with and managing the unfolding AI process and more ways of understanding the AI model within a performance.
Over three different performances, this particular setup started to feel like a more comfortable and expressive instrument with potential to sit within and drive different improvisational scenarios.
It's worth noting that at this point, I was more likely to perform with a Raspberry Pi 4 or 5 rather than the Zero 2 W used in early prototypes. 
Both the boot and inference times were faster on the full-size Raspberry Pis, a practical and musical consideration for performance.

The S-1/QuNeo intelligent setup was used in the final performance experience listed in Table~\ref{tab:perf-corpus}. 
This was, notably, an improvisation festival where I used the same instrument in three performances with different groups of musicians including acoustic instruments, acoustic instruments with electronic effects and augmentations, and electronic musical instruments.
The challenge here was to manage the, sometimes unruly, intelligent musical instrument in an environment where the \emph{vibe} of a particular performance might be unknown in advance and range from sensitive sparse textures to a full-on wall of noise. With the visual feedback and more flexible controllers in the intelligent setup, this finally felt achievable.

\begin{figure}[t]
	\centering
	\includegraphics[width=0.99\linewidth]{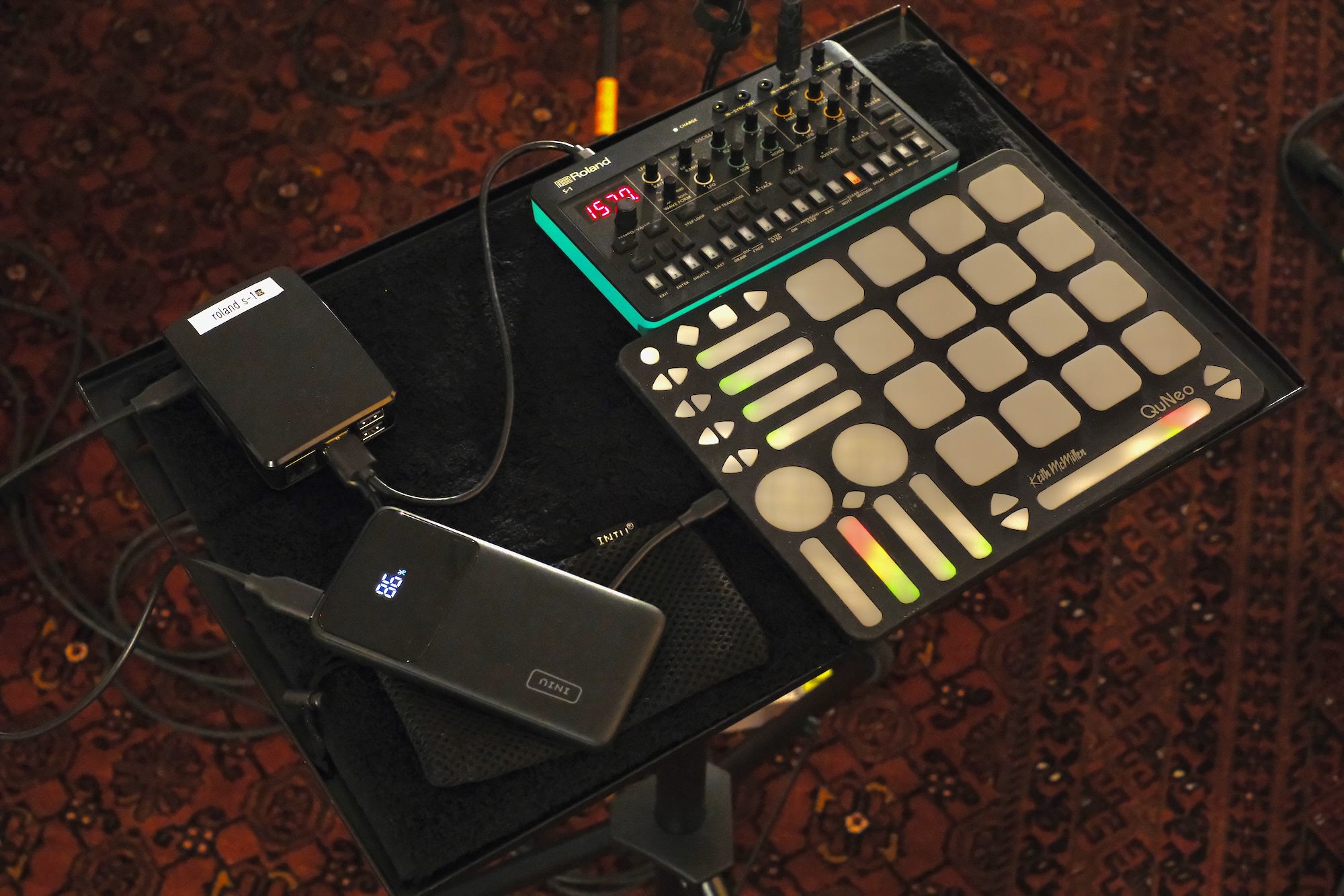}
	\caption{An intelligent setup of the Roland S-1 and Keith McMillen QuNeo interfaces with a Raspberry Pi 5 as used in festival performances in January 2026}
    \label{fig:intelligent-setup}
\end{figure}

\begin{table*}[t]
\caption{Configuration of the performances, recordings, and demonstrations considered as part of the first-person artistic research process in this work throughout 2024--2026. This list does not include rehearsal or practice sessions that also took place.
A "*" in the configuration column denotes where more than one performer used an intelligent instrument studied in this paper, other duo and group performances involved one intelligent instrument (performed by the author) with other performers using acoustic and conventional electronic instruments.}
\label{tab:perf-corpus}
\begin{tabular}{@{}llll@{}}
Date           & Type           & Configuration & Intelligent Instrument      \\ \midrule
February 2024  & Recording      & Solo          & Volca                       \\
May 2024       & Recording      & Solo          & MicroFreak                  \\
May 2024       & Performance    & Group*        & MicroFreak, DAW (AUM, Live) \\
June 2024      & Performance    & Duo           & MicroFreak                  \\
June 2024      & Performance    & Group         & MicroFreak                  \\
September 2024 & Performance    & Solo          & MicroFreak, S-1             \\
September 2024 & Performance    & Duo           & MicroFreak                  \\
October 2024   & Performance    & Group         & S-1                         \\
December 2024  & Recording      & Duo*          & DAW (AUM, Live)             \\
January 2025   & Recording      & Solo          & S-1, DAW                    \\
August 2025    & Performance    & Group         & S-1, DAW                    \\
November 2025  & Performance    & Duo           & Setup (S-1/xTouch)          \\
November 2025  & Performance    & Duo           & Setup (S-1/QuNeo)           \\
December 2025  & Recording      & Duo           & Setup (S-1/QuNeo)           \\
January 2026   & Performance    & Group         & Setup (S-1/QuNeo)           \\
\end{tabular}
\end{table*}

\section{Discussion}

This work has described a two-year practice of intelligent instrument building and exploration in performances and recordings. The goal of this process was to expand the design space of intelligent musical instruments by developing and evolving many prototypes. 
In this section, I seek to distill the development and performance experiences into design implications relating to how generative AI models may be used in a productive way for intelligent instrument design.

\subsection{Mappings and Flexible Configuration}

A main finding in this research was that expanding the mapping capability of the generative AI platform, that is enabling changes in how the AI model is connected to physical controls and instrument parameters, paid dividends in terms of different instrument designs. 
This reflects on calls to expand interactivity when operating with ML models~\cite{jourdan_nime_ml_review}. 
Connecting the generative AI outputs to simultaneously control multiple timbral parameters led to new musical potential in the intelligent MicroFreak. Adapting the platform to a DAW exposed a convenient way to play with how AI inputs and outputs could affect live performance. Supporting simultaneous IO from multiple devices in intelligent setups expanded the gestural affordances for managing and tracking AI behaviours.

Adjusting the mapping of control parameters to and from a generative AI model allowed me to find where a model could have the best musical impact within an intelligent instrument. 
The earlier prototypes demonstrated that the AI actions were more usefully focussed on what a human could not do (e.g., turn five knobs at once) rather than focussing on note production. 
Mapping the AI outputs to a visual indicator turned out to support better real-time understanding.

This research suggests that re-mapping may be an alternative to retraining where training might be expensive or inconvenient. 
Early small-data ML models~\cite{fiebrink_years_ml} can be retrained within seconds, while the models in this project take minutes to hours, and audio generation models may take days~\cite{stefansdottir_intelligent_2025}. 
Supporting remapping before retraining led to a faster and environmentally aware~\cite{sustainable-internet-of-musical-things} prototyping process.

\paragraph{Design Implications:} Flexible (re)mapping enables discovery of productive AI interactions. AI actions should be focussed on what a musician cannot do and visual feedback can be critical.

\subsection{AI Models as a Transportable Instrument Component}

The generative AI model in this work was trained entirely on self-produced data using standard computer systems, a clear contrast with the industrial-scale machine learning approach and its many ethical problems. This connects well with Vigliensoni's small data mindset~\cite{Vigliensoni:2022} and challenges the assumption that AI necessarily requires unethically sourced data and unsustainable computational resources.
A finding of this research was that one trained generative AI model turned out to be useful across a range of intelligent musical instrument prototypes.
While other small-data music models have been published~\cite[e.g.,][]{stefansdottir_intelligent_2025}, this paper demonstrates the utility that this approach can deliver.
The trained model itself became a design component similar to the many interchangeable devices, modules and interfaces that electronic musicians use to assemble new interfaces. 
The artistic process revealed new uses for the same model and the mapping functionality extended its utility into applications quite disconnected from the originally collected data. 
Through the process of performance and practice I got to know this model better and found better ways to apply it.
This speaks to sustainability~\cite{sustainable-internet-of-musical-things}. 
A useful generative model can, like a trusted effects pedal or a filter module, be a transportable component within an electronic music practice.

\paragraph{Design Implications:} Small-data generative AI should be applied in individual creative practices. Models can become a transportable design resource for intelligent musical instruments.

\subsection{Co-Creative Performance}

Throughout this research, the interaction strategy between AI and human input has used a typical call-and-response mode not dissimilar from previous instruments such as Continuator~\cite{Pachet:2003wd} or EMPI~\cite{martin_empi}. The idea of very fast switchover between human and AI control of an instrument in this research appears to be novel, contrasting with recent approaches using footswitches to alternate AI listening and generation~\cite{living_looper_shepardson}, and to produce a unique interactive experience. 
Rather than sharing a musical process with a separate agent, the intelligent instrument feels like a continually changing device, like a free-running oscillator, feedback system, or sequencer, that a performer can guide and adjust, but not fully control.
The ability to instantly take over control through natural performance gestures helped address the challenge of guiding or \emph{rescuing}~\cite{stefansdottir_intelligent_2025} intelligent instruments away from unwanted sounds.
The idea of shared control has been previously studied, but present approaches in LLM-based AI tend towards complete control of creation with only high-level management through prompting.
This research demonstrates that new kinds of co-creative experiences are still possible, musically productive, and fun.

\paragraph{Design Implication:}
Very fast interleaving between AI and human inputs produces a collaborative, productive, and fun musical experience.

\subsection{Accessible Intelligent Instruments}

An early motivation for this project was the discovery that the cheapest Raspberry Pi Zero 2 W (15USD) might be able to run small generative AI models, addressing cost issues as an accessibility barrier~\cite{longevity-of-bespoke} for intelligent instrument design. The Intelligent Volca, MicroFreak, and S-1 confirmed that the Zero 2 W worked in different performance situations while revealing the limitations in terms of boot time when using this very low-end hardware. An advantage of the cheap Raspberry Pis was that I could set up multiple for use in group performances, loaning to collaborators, or dedicate to different instruments. 
The idea of dedicating single-board computers to different interfaces was established with Satellite CCRMA~\cite{satelliteCCRMA2011}, but it's certainly easier to do this in practice with the cheapest hardware. Of course, I ended up using the more expensive Raspberry Pi 4 and 5 in my personal creative practice, prioritising speed.

While low costs help to make intelligent instruments more inclusive~\cite{jourdan_nime_ml_review}, a focus on the ``new'' raises questions of sustainability~\cite{sustainable-internet-of-musical-things, the_o_in_nime}. 
Taking advantage of MIDI as the main communication standard in this project accelerated opportunities to prototype new designs but also sustainably reuse and retrofit existing instruments with a generative AI model.
It was easier to prototype intelligent instruments by connecting existing hardware to an existing model in a new way, putting the focus on whether the interaction is musically satisfying.
This approach built on previous use of OSC~\cite{Martin2019} but required more careful configuration approach to define mappings.
It's notable that MIDI was particularly useful for connecting the system to a computer for the Intelligent DAW where existing MIDI mapping functionality in Live and AUM was available.

\paragraph{Design implication:} Using cheap single board computers and standard MIDI lowers barriers to sustainable and inclusive prototyping.

\subsection{Limitations}

This research has explored intelligent instrument performance from a first-person perspective. 
This approach is appropriate for understanding an exploratory design process over a long period of time (2 years).
In future, it would be beneficial to extend this research to explore intelligent instrument designs from other creative practitioners.
Reflecting on instruments from musicians of different backgrounds would likely lead to additional design implications.
In this research, the potential to update the AI model over multiple practice and performance sessions was not explored in a systematic way. Again, the evolution of a small-data AI model could be an additional aspect to the intelligent instrument design space.

\section{Conclusions}\label{conclusions}

This paper has presented a new generative AI platform for designing and experimenting with new intelligent musical instruments. 
I have argued that the design space for intelligent musical instruments using present AI techniques is still only beginning to be explored and that small-data approaches, such as this system, can lead to artist-centred musical AI.
Experience using this platform in five prototype instruments over two years of live performances and recordings has shown that (re)mapping can replace retraining for discovering AI interaction, that fast input interleaving can be a co-creative strategy, that small-data AI models can be a transportable design resource, and that cheap hardware can lower barriers to inclusion.
In the hardware synth examples, AI-control over notes and timbral parameters led to unique performances and sounds, making up for limitations in synth interface design.
The intelligent DAW demonstrated the flexibility of the system when coupled with highly configurable DAW MIDI mappings, expanding the design possibilities for interactions with a DAW as an instrument.
The intelligent setup allowed combinations of hardware interfaces to open up multiple co-creation styles with the AI system.
This work contributes design possibilities explored in one creative practice with this platform through a first-person reflective process to the ongoing conversation within the computer music and broader HCI communities about how generative AI might be applied in interactive systems.
Future work could examine the impact of updating models over time with artist-sourced data, and engage artists in co-design processes for creating new intelligent musical instruments.

\section{Ethical Standards}

This paper involved autobiographical design and study of new interfaces from the author's perspective.
Where performances were in group settings, other group members were involved as creative collaborators rather than research participants and have been acknowledged below.
The machine learning models used in this research were created with the author's own data.

\begin{acks}
My thanks to collaborators and creative partners who made music with me as part of this project: Alec Hunter, Yichen Wang, Sandy Ma, Richard Johnson and the SoundOut Collective.
\end{acks}

\bibliographystyle{ACM-Reference-Format}
\bibliography{references} 

@inproceedings{bowers-room-feedback,
 address = {Mexico City, Mexico},
 articleno = {71},
 author = {John M Bowers},
 booktitle = {Proceedings of the International Conference on New Interfaces for Musical Expression},
 doi = {10.5281/zenodo.11189266},
 editor = {Miguel Ortiz and Adnan Marquez-Borbon},
 issn = {2220-4806},
 month = {May},
 numpages = {10},
 pages = {511--520},
 title = {A Hapless But Entertaining Roar: Developing a Room Feedback System through Artistic Research and Aesthetic Reflection},
 track = {Papers},
 url = {http://nime.org/proceedings/2023/nime2023_71.pdf},
 year = {2023}
}

@inproceedings{the_o_in_nime,
 address = {Mexico City, Mexico},
 articleno = {14},
 author = {Raul Masu and Fabio Morreale and Alexander Refsum Jensenius},
 booktitle = {Proceedings of the International Conference on New Interfaces for Musical Expression},
 doi = {10.5281/zenodo.11189120},
 editor = {Miguel Ortiz and Adnan Marquez-Borbon},
 issn = {2220-4806},
 month = {May},
 numpages = {10},
 pages = {106--115},
 title = {The O in NIME: Reflecting on the Importance of Reusing and Repurposing Old Musical Instruments},
 track = {Papers},
 url = {http://nime.org/proceedings/2023/nime2023_14.pdf},
 year = {2023}
}

@inproceedings{longevity-of-bespoke,
 address = {Birmingham, UK},
 author = {Lucas, Alex and Ortiz, Miguel and Schroeder, Franziska},
 booktitle = {Proceedings of the International Conference on New Interfaces for Musical Expression},
 doi = {10.5281/zenodo.4813338},
 editor = {Romain Michon and Franziska Schroeder},
 issn = {2220-4806},
 month = {July},
 pages = {243--248},
 presentation-video = {https://youtu.be/cLguyuZ9weI},
 publisher = {Birmingham City University},
 title = {The Longevity of Bespoke, Accessible Music Technology: A Case for Community},
 url = {https://www.nime.org/proceedings/2020/nime2020_paper46.pdf},
 year = {2020}
}

@ARTICLE{sustainable-internet-of-musical-things,
  author={Masu, Raul and Merendino, Nicolò and Rodã, Antonio and Turchet, Luca},
  journal={IEEE Access}, 
  title={Sustainable Internet of Musical Things: Strategies to Account for Environmental and Social Sustainability in Network-Based Interactive Music Systems}, 
  year={2024},
  volume={12},
  number={},
  pages={62818-62833},
  doi={10.1109/ACCESS.2024.3393468}}

@inproceedings{living_looper_shepardson,
 address = {Canberra, Australia},
 articleno = {36},
 author = {Victor Shepardson and Halla Steinunn Stefánsdóttir and Thor Magnusson},
 booktitle = {Proceedings of the International Conference on New Interfaces for Musical Expression},
 doi = {10.5281/zenodo.15698851},
 editor = {Doga Cavdir and Florent Berthaut},
 issn = {2220-4806},
 month = {June},
 numpages = {6},
 pages = {255--260},
 title = {Evolving the Living Looper: Artistic Research, Online Learning, and Tentacle Pendula},
 track = {Paper},
 url = {http://nime.org/proceedings/2025/nime2025_36.pdf},
 year = {2025}
}

@article{stefansdottir_intelligent_2025,
  author   = {Stefánsdóttir, Halla Steinunn  and Magnusson, Thor },
  title    = {Of altered instrumental relations: a practice-led inquiry into agency through musical performance with neural audio synthesis and violin},
  journal  = {Frontiers in Computer Science},
  volume   = {Volume 7 - 2025},
  year     = {2025},
  url      = {https://www.frontiersin.org/journals/computer-science/articles/10.3389/fcomp.2025.1578595},
  doi      = {10.3389/fcomp.2025.1578595},
  issn     = {2624-9898}
}

@misc{MIDI-Manufacturers-Association:2014aa,
	author = {{MIDI Manufacturers Association}},
	title = {{MIDI} 1.0 Electrical Specification Update},
	url = {https://www.midi.org/specifications-old/item/midi-din-electrical-specification},
	year = {2014}}

@inproceedings{Naess:2019aa,
	author = {N{\ae}ss, Torgrim Rudland and Martin, Charles Patrick},
	booktitle = {Proceedings of the International Conference on New Interfaces for Musical Expression},
	pages = {79--82},
	title = {A Physical Intelligent Instrument using Recurrent Neural Networks},
        doi = {10.5281/zenodo.3672874},
	year = {2019}}

@inproceedings{Martin2019,
 author = {Charles Patrick Martin and Jim Torresen},
 booktitle = {Proceedings of the International Conference on New Interfaces for Musical Expression},
 doi = {10.5281/zenodo.3672952},
 pages = {260--265},
 title = {An Interactive Musical Prediction System with Mixture Density Recurrent Neural Networks},
 year = {2019}
}

@software{impsy_software_zenodo,
  author       = {Charles Patrick Martin},
  title        = {IMPSY: An Intelligent Musical Instrument Platform},
  year         = 2026,
  publisher    = {Zenodo},
  version      = {v0.7.0-alpha},
  doi          = {10.5281/zenodo.2580175},
  url          = {https://github.com/cpmpercussion/impsy/},
}

@article{martin_empi,
	author = {Martin, Charles Patrick and Glette, Kyrre and Nygaard, T{\o}nnes Frostad and Torresen, Jim},
	doi = {10.3389/frai.2020.00006},
	issn = {2624-8212},
	journal = {Frontiers in Artificial Intelligence},
	title = {Understanding Musical Predictions With an Embodied Interface for Musical Machine Learning},
	volume = {3},
	year = {2020}}

@software{keras-mdn-layer,
	author = {Charles Martin},
	doi = {10.5281/zenodo.1482347},
	month = nov,
	publisher = {Zenodo},
	title = {cpmpercussion/keras-mdn-layer},
	url = {https://doi.org/10.5281/zenodo.1482347},
	version = {v0.3.0},
	year = 2019}

@inproceedings{pelinski_pipeline_bela,
 author = {Teresa Pelinski and Rodrigo Diaz and Adan L. Benito Temprano and Andrew McPherson},
 booktitle = {Proceedings of the International Conference on New Interfaces for Musical Expression},
 pages = {160--166},
 title = {Pipeline for recording datasets and running neural networks on the Bela embedded hardware platform},
 url = {http://nime.org/proceedings/2023/nime2023_22.pdf},
 year = {2023}
}

@inproceedings{mcpherson-bela-2016,
 address = {Brisbane, Australia},
 author = {Andrew McPherson and Robert Jack and Giulio Moro},
 booktitle = {Proceedings of the International Conference on New Interfaces for Musical Expression},
 doi = {10.5281/zenodo.3964611},
 isbn = {978-1-925455-13-7},
 issn = {2220-4806},
 pages = {20--25},
 publisher = {Queensland Conservatorium Griffith University},
 title = {Action-Sound Latency: Are Our Tools Fast Enough?},
 track = {Papers},
 url = {http://www.nime.org/proceedings/2016/nime2016_paper0005.pdf},
 year = {2016}
}

@inproceedings{jourdan_nime_ml_review,
 author = {Théo Jourdan and Baptiste Caramiaux},
 booktitle = {Proceedings of the International Conference on New Interfaces for Musical Expression},
 numpages = {13},
 title = {Machine Learning for Musical Expression: A Systematic Literature Review},
 url = {http://nime.org/proceedings/2023/nime2023_46.pdf},
 year = {2023}
}

@incollection{visi_iml_gesture,
author="Visi, Federico Ghelli
and Tanaka, Atau",
title="Interactive Machine Learning of Musical Gesture",
bookTitle="Handbook of Artificial Intelligence for Music: Foundations, Advanced Approaches, and Developments for Creativity",
year="2021",
publisher="Springer",
pages="771--798",
isbn="978-3-030-72116-9",
doi="10.1007/978-3-030-72116-9_27",
}

@article{caillon_rave,
  author       = {Antoine Caillon and
                  Philippe Esling},
  title        = {{RAVE:} {A} variational autoencoder for fast and high-quality neural
                  audio synthesis},
  journal      = {CoRR},
  volume       = {abs/2111.05011},
  year         = {2021},
  eprinttype    = {arXiv},
  eprint       = {2111.05011},
}

@inproceedings{Vigliensoni:2022,
	title = {A small-data mindset for generative AI creative work},
	author = {Gabriel Vigliensoni and Phoenix Perry and Rebecca Fiebrink},
	booktitle = {Proceedings of the CHI 2022 Workshop on Generative AI and HCI},
	year = {2022},
	url = {https://doi.org/10.5281/zenodo.7086327},
	doi = {10.5281/zenodo.7086327}
}

@inproceedings{teachablemachine,
author = {Carney, Michelle and Webster, Barron and Alvarado, Irene and Phillips, Kyle and Howell, Noura and Griffith, Jordan and Jongejan, Jonas and Pitaru, Amit and Chen, Alexander},
title = {Teachable Machine: Approachable Web-Based Tool for Exploring Machine Learning Classification},
year = {2020},
publisher = {ACM},
address = {New York, NY, USA},
doi = {10.1145/3334480.3382839},
booktitle = {Extended Abstracts of the 2020 CHI Conference on Human Factors in Computing Systems},
pages = {1–8},
}

@article{musicpublishers,
title = {Music publishers sue Amazon-backed AI company over song lyrics},
author = {Blake Montgomery},
year = {2023},
journal = {The Guardian},
url = {https://www.theguardian.com/technology/2023/oct/19/music-lawsuit-ai-song-lyrics-anthropic},
month = {oct}
}

@inproceedings{fiebrink_years_ml,
 author = {Fiebrink, Rebecca and Sonami, Laetitia},
 booktitle = {Proceedings of the International Conference on New Interfaces for Musical Expression},
 doi = {10.5281/zenodo.4813334},
 pages = {237--242},
 title = {Reflections on Eight Years of Instrument Creation with Machine Learning},
 year = {2020}
}

@inproceedings{fiebrink_meta-instrument_2009,
	title = {A {Meta}-{Instrument} for {Interactive}, {On}-the-fly {Machine} {Learning}},
	booktitle = {Proc. {NIME}},
	author = {Fiebrink, Rebecca and Trueman, Dan and Cook, Perry R},
	year = {2009},
	pages = {280--285},
}

@article{ji_survey_2024,
	title = {A {Survey} on {Deep} {Learning} for {Symbolic} {Music} {Generation}: {Representations}, {Algorithms}, {Evaluations}, and {Challenges}},
	volume = {56},
	doi = {10.1145/3597493},
	number = {1},
	journal = {ACM Computing Surveys},
	author = {Ji, Shulei and Yang, Xinyu and Luo, Jing},
	year = {2024},
	pages = {1--39},
}

@article{deruty_development_2022,
	title = {On the {Development} and {Practice} of {AI} {Technology} for {Contemporary} {Popular} {Music} {Production}},
	volume = {5},
	issn = {2514-3298},
	url = {https://transactions.ismir.net/article/10.5334/tismir.100/},
	number = {1},
	urldate = {2024-01-19},
	journal = {Transactions of the International Society for Music Information Retrieval},
	author = {Deruty, Emmanuel and Grachten, Maarten and Lattner, Stefan and Nistal, Javier and Aouameur, Cyran},
	month = feb,
	year = {2022},
	pages = {35},
}

@inproceedings{autoethnography-hci,
author = {Kaltenhauser, Annika and Stefanidi, Evropi and Sch\"{o}ning, Johannes},
title = {Playing with Perspectives and Unveiling the Autoethnographic Kaleidoscope in HCI – A Literature Review of Autoethnographies},
year = {2024},
isbn = {9798400703300},
publisher = {Association for Computing Machinery},
address = {New York, NY, USA},
url = {https://doi.org/10.1145/3613904.3642355},
doi = {10.1145/3613904.3642355},
booktitle = {Proceedings of the 2024 CHI Conference on Human Factors in Computing Systems},
articleno = {819},
numpages = {20},
location = {Honolulu, HI, USA},
series = {CHI '24}
}

@inproceedings{autobiographical-design,
author = {Neustaedter, Carman and Sengers, Phoebe},
title = {Autobiographical design in HCI research: designing and learning through use-it-yourself},
year = {2012},
isbn = {9781450312103},
publisher = {Association for Computing Machinery},
address = {New York, NY, USA},
doi = {10.1145/2317956.2318034},
booktitle = {Proceedings of the Designing Interactive Systems Conference},
pages = {514–523},
numpages = {10},
location = {Newcastle Upon Tyne, United Kingdom},
series = {DIS '12}
}

@incollection{Smith:2009gf,
	author = {Hazel Smith and Roger T. Dean},
	booktitle = {Practice-led Research, Research-led Practice in the Creative Arts},
	editor = {Hazel Smith and Roger T. Dean},
	pages = {1--38},
	publisher = {Edinburgh University Press},
	title = {Introduction: Practice-led Research, Research-led Practice --- Towards the Iterative Cyclic Web},
	year = {2009}}

@inproceedings{artistic-narratives-hci,
author = {Sturdee, Miriam and Lewis, Makayla and Strohmayer, Angelika and Spiel, Katta and Koulidou, Nantia and Alaoui, Sarah Fdili and Urban Davis, Josh},
title = {A Plurality of Practices: Artistic Narratives in HCI Research},
year = {2021},
isbn = {9781450383769},
publisher = {Association for Computing Machinery},
address = {New York, NY, USA},
url = {https://doi.org/10.1145/3450741.3466771},
doi = {10.1145/3450741.3466771},
booktitle = {Proceedings of the 13th Conference on Creativity and Cognition},
articleno = {35},
numpages = {14},
location = {Virtual Event, Italy},
series = {C\&C '21}
}

@inproceedings{genaichi2022,
author = {Muller, Michael and Chilton, Lydia B and Kantosalo, Anna and Martin, Charles Patrick and Walsh, Greg},
title = {GenAICHI: Generative AI and HCI},
year = {2022},
isbn = {9781450391566},
publisher = {Association for Computing Machinery},
address = {New York, NY, USA},
url = {https://doi.org/10.1145/3491101.3503719},
doi = {10.1145/3491101.3503719},
booktitle = {Extended Abstracts of the 2022 CHI Conference on Human Factors in Computing Systems},
articleno = {110},
numpages = {7},
location = {New Orleans, LA, USA},
series = {CHI EA '22}
}

@techreport{goldmedia_ai_2024,
	title = {{AI} and {Music}: {Market} {Development} of {AI} in the {Music} {Sector} and {Impact} on {Music} {Creators} in {Australia} and {New} {Zealand}},
	url = {https://www.apraamcos.com.au/about/supporting-the-industry/research-papers/aiandmusic},
	author = {Goldmedia},
	month = aug,
	year = {2024},
}

@inproceedings{ethical_genaudio_2023,
author = {Barnett, Julia},
title = {The Ethical Implications of Generative Audio Models: A Systematic Literature Review},
year = {2023},
isbn = {9798400702310},
publisher = {Association for Computing Machinery},
address = {New York, NY, USA},
url = {https://doi.org/10.1145/3600211.3604686},
doi = {10.1145/3600211.3604686},
booktitle = {Proceedings of the 2023 AAAI/ACM Conference on AI, Ethics, and Society},
pages = {146–161},
numpages = {16},
location = {Montr\'{e}al, QC, Canada},
series = {AIES '23}
}

@inproceedings{sound_design_ai_2024,
author = {Kamath, Purnima and Morreale, Fabio and Bagaskara, Priambudi Lintang and Wei, Yize and Nanayakkara, Suranga},
title = {Sound Designer-Generative AI Interactions: Towards Designing Creative Support Tools for Professional Sound Designers},
year = {2024},
isbn = {9798400703300},
publisher = {Association for Computing Machinery},
address = {New York, NY, USA},
url = {https://doi.org/10.1145/3613904.3642040},
doi = {10.1145/3613904.3642040},
booktitle = {Proceedings of the 2024 CHI Conference on Human Factors in Computing Systems},
articleno = {730},
numpages = {17},
location = {Honolulu, HI, USA},
series = {CHI '24}
}

@article{Lewis:2000fu,
	author = {Lewis, George E.},
	doi = {10.1162/096112100570585},
	journal = {Leonardo Music Journal},
	pages = {33--39},
	title = {Too Many Notes: Computers, Complexity and Culture in ``{V}oyager''},
	volume = {10},
	year = {2000}
}

@article{Pachet:2003wd,
	author = {Fran{\c c}ois Pachet},
	doi = {10.1076/jnmr.32.3.333.16861},
	journal = {Journal of New Music Research},
	number = {3},
	pages = {333--341},
	title = {The Continuator: Musical Interaction With Style},
	volume = {32},
	year = {2003}
}

@article{roads-music-AI-1985,
author = {Roads, Curtis},
title = {Research in music and artificial intelligence},
year = {1985},
issue_date = {June 1985},
publisher = {Association for Computing Machinery},
address = {New York, NY, USA},
volume = {17},
number = {2},
issn = {0360-0300},
url = {https://doi.org/10.1145/4468.4469},
doi = {10.1145/4468.4469},
journal = {ACM Comput. Surv.},
month = jun,
pages = {163–190},
numpages = {28}
}

@incollection{Biles:2007aa,
	address = {London},
	author = {John A. Biles},
	booktitle = {Evolutionary Computer Music},
	doi = {10.1007/978-1-84628-600-1_7},
	editor = {Miranda, Eduardo Reck and Biles, John Al},
	pages = {137--169},
	publisher = {Springer London},
	title = {Improvizing with Genetic Algorithms: GenJam},
	year = {2007}
}

@inproceedings{tahiroglu-aiterity-2021,
 address = {Shanghai, China},
 articleno = {80},
 author = {Tahiroğlu, Koray and Kastemaa, Miranda and Koli, Oskar},
 booktitle = {Proceedings of the International Conference on New Interfaces for Musical Expression},
 doi = {10.21428/92fbeb44.3d0e9e12},
 issn = {2220-4806},
 month = {June},
 title = {AI-terity 2.0: An Autonomous NIME Featuring GANSpaceSynth Deep Learning Model},
 year = {2021}
}

@incollection{tahirouglu-evolving-2022,
  title={Evolving musical expectations: Mutual correlation between a human musician and an AI musical instrument},
  author={Tahiro{\u{g}}lu, Koray},
  booktitle={Embodied gestures},
  editors={Tomás Calderón, Enrique and Gorbach, Thomas and Tellioğlu, Hilda and Kaltenbrunner, Martin},
  doi={10.34727/2022/isbn.978-3-85448-047-1_4},
  pages={30--36},
  year={2022},
  publisher={Technische Universit{\"a}t Wien}
}

@inproceedings{satelliteCCRMA2011,
	address = {Oslo, Norway},
	author = {Edgar Berdahl and Wendy Ju},
	booktitle = {Proceedings of the International Conference on New Interfaces for Musical Expression},
	editor = {Alexander Refsum Jensenius and Anders Tveit and Rolf Inge God{\o}y and Dan Overholt},
	numpages = {6},
	pages = {173--178},
	publisher = {University of Oslo},
	series = {NIME '11},
	title = {Satellite {CCRMA}: A Musical Interaction and Sound Synthesis Platform},
	url = {http://www.nime.org/proceedings/2011/nime2011_173.pdf},
	year = {2011}
}

@inproceedings{Berdahl:2013aa,
	address = {Daejeon, Republic of Korea},
	author = {Edgar Berdahl and Spencer Salazar and Myles Borins},
	booktitle = {Proceedings of the International Conference on New Interfaces for Musical Expression},
	editor = {Woon Seung Yeo and Kyogu Lee and Alexander Sigman and Haru Ji and Graham Wakefield},
	pages = {325--330},
	publisher = {KAIST},
	series = {NIME '13},
	title = {Embedded Networking and Hardware-Accelerated Graphics with Satellite CCRMA},
	year = {2013}}

@inproceedings{visi-sophtar-2024,
 address = {Utrecht, Netherlands},
 articleno = {22},
 author = {Federico Visi},
 booktitle = {Proceedings of the International Conference on New Interfaces for Musical Expression},
 doi = {10.5281/zenodo.13904810},
 editor = {S M Astrid Bin and Courtney N. Reed},
 issn = {2220-4806},
 month = {September},
 numpages = {7},
 pages = {142--148},
 presentation-video = {},
 title = {The Sophtar: a networkable feedback string instrument with embedded machine learning},
 track = {Papers},
 url = {http://nime.org/proceedings/2024/nime2024_22.pdf},
 year = {2024}
}

@inproceedings{privato-hauntography-2024,
 address = {Utrecht, Netherlands},
 articleno = {63},
 author = {Nicola Privato and Thor Magnusson},
 booktitle = {Proceedings of the International Conference on New Interfaces for Musical Expression},
 doi = {10.5281/zenodo.13904901},
 editor = {S M Astrid Bin and Courtney N. Reed},
 issn = {2220-4806},
 month = {September},
 numpages = {7},
 pages = {432--438},
 presentation-video = {},
 title = {Querying the Ghost: AI Hauntography in NIME},
 track = {Papers},
 url = {http://nime.org/proceedings/2024/nime2024_63.pdf},
 year = {2024}
}

@article{daws-reuter-2021,
author = {Anders Reuter},
title = {Who let the DAWs Out? The Digital in a New Generation of the Digital Audio Workstation},
journal = {Popular Music and Society},
volume = {45},
number = {2},
pages = {113--128},
year = {2022},
publisher = {Routledge},
doi = {10.1080/03007766.2021.1972701}
}

@inproceedings{magenta-studio-2019,
title	= {Magenta Studio: Augmenting Creativity with Deep Learning in Ableton Live},
author	= {Adam Roberts and Jesse Engel and Yotam Mann and Jon Gillick and Claire Kayacik and Signe Nørly and Monica Dinculescu and Carey Radebaugh and Curtis Hawthorne and Douglas Eck},
year	= {2019},
URL	= {http://musicalmetacreation.org/buddydrive/file/mume_2019_paper_2/},
booktitle	= {Proceedings of the International Workshop on Musical Metacreation (MUME)}
}

@article{taxonomy-music-interactions-parkinson-2025, 
title={Improvising with Machines: A taxonomy of musical interactions}, 
DOI={10.1017/S1355771824000268}, 
journal={Organised Sound}, 
author={Parkinson, Adam and Dunning, Graham}, 
year={2025}, 
pages={1–12}
}

@incollection{Clauhs:2023aa,
	author = {Matthew Clauhs and Brian Dozoretz},
	booktitle = {The Routledge Companion to Creativities in Music Education},
    doi = {10.4324/9781003248194},
	editor = {Clint Randles and Pamela Burnard},
	publisher = {Taylor \& Francis Group},
	title = {The DAW Revolution},
	year = {2023}
}

@incollection{Davis:2022,
    author = {Dylan Davis},
    title = {Compute and Resonate: An Ongoing Experiment in Creating Acid Music Using Accessible Artificial Intelligence and Computer-Based Generative Tools},
    booktitle = {Designing Interactions for Music and Sound},
    editor = {Michael Filimowicz},
    publisher = {Routledge},
    year = {2022} 
}

@inproceedings{design-principles-genai-2024,
author = {Weisz, Justin D. and He, Jessica and Muller, Michael and Hoefer, Gabriela and Miles, Rachel and Geyer, Werner},
title = {Design Principles for Generative AI Applications},
year = {2024},
isbn = {9798400703300},
publisher = {Association for Computing Machinery},
address = {New York, NY, USA},
url = {https://doi.org/10.1145/3613904.3642466},
doi = {10.1145/3613904.3642466},
booktitle = {Proceedings of the 2024 CHI Conference on Human Factors in Computing Systems},
articleno = {378},
numpages = {22},
location = {Honolulu, HI, USA},
series = {CHI '24}
}

@inproceedings{improvisation-arts-hci-2018,
author = {Kang, Laewoo (Leo) and Jackson, Steven J. and Sengers, Phoebe},
title = {Intermodulation: Improvisation and Collaborative Art Practice for HCI},
year = {2018},
isbn = {9781450356206},
publisher = {Association for Computing Machinery},
address = {New York, NY, USA},
url = {https://doi.org/10.1145/3173574.3173734},
doi = {10.1145/3173574.3173734},
booktitle = {Proceedings of the 2018 CHI Conference on Human Factors in Computing Systems},
pages = {1–13},
numpages = {13},
location = {Montreal QC, Canada},
series = {CHI '18}
}

@inproceedings{state-of-arts-inchi-2022,
author = {Sturdee, Miriam and Lewis, Makayla and Gamboa, Mafalda and Hoang, Thuong and Miers, John and \v{S}morgun, Ilja and Jain, Pranjal and Strohmayer, Angelika and Fdili Alaoui, Sarah and Wodtke, Christina R},
title = {The State of the (CHI)Art},
year = {2022},
isbn = {9781450391566},
publisher = {Association for Computing Machinery},
address = {New York, NY, USA},
url = {https://doi.org/10.1145/3491101.3503722},
doi = {10.1145/3491101.3503722},
booktitle = {Extended Abstracts of the 2022 CHI Conference on Human Factors in Computing Systems},
articleno = {104},
numpages = {6},
location = {New Orleans, LA, USA},
series = {CHI EA '22}
}

@article{interactive-art-hci-2019,
title = {From rituals to magic: Interactive art and HCI of the past, present, and future},
journal = {International Journal of Human-Computer Studies},
volume = {131},
pages = {108-119},
year = {2019},
note = {50 years of the International Journal of Human-Computer Studies. Reflections on the past, present and future of human-centred technologies},
issn = {1071-5819},
doi = {https://doi.org/10.1016/j.ijhcs.2019.06.005},
url = {https://www.sciencedirect.com/science/article/pii/S1071581919300758},
author = {Myounghoon Jeon and Rebecca Fiebrink and Ernest A. Edmonds and Damith Herath}
}

\end{document}